# Throughput Optimal Uplink Scheduling in Heterogeneous PLC and LTE Communication for Delay Aware Smart Grid Applications

Qiyue LI[1,2], Tengfei CAO[1,2], Wei SUN[1,2], and Weitao LI[1,2]

[1] Hefei University of Technology, School of Electrical Engineering and Automation, Hefei 230009, Anhui, China
[2] Engineering Technology Research Center of Industrial Automation, Hefei 230009, Anhui, China
liqiyue@mail.ustc.edu.cn
tfcao@mail.hfut.edu.cn
wsun@hfut.edu.cn
wtli@hfut.edu.cn

**Abstract.**
Smart grid is an energy network that integrates advanced power equipment, communication technology and control technology. It can transmit two-way power and data among all components of the grid at the same time. The existing smart grid communication technologies include power line carrier (PLC) communication, industrial Ethernet, passive optical networks and wireless communication, each of which have different advantages. Due to the complex application scenarios, massive sampling points and high transmission reliability requirements, a single communication method cannot fully meet the communication requirements of smart grid, and heterogeneous communication modes are required. In addition, with the development of cellular technology, long term evolution (LTE)-based standards have been identified as a promising technology that can meet the strict requirements of various operations in smart grid. In this paper, we analyze the advantages and disadvantages of PLC and LTE communication, and design a network framework for PLC and LTE communication uplink heterogeneous communication in smart grid. Then, we propose an uplink scheduling transmission method for sampling data with optimized throughput according to the requirements of system delay and reliability. Then, we use the formula derivation to prove the stability and solvability of the scheduling system in theory. Finally, the simulation results show that under the condition of satisfying the delay requirement, our proposed framework can optimally allocate the wireless communication resource and maximize the throughput of the uplink transmission system.

## 1 Introduction

Smart grid is built on the basis of an integrated high-speed two-way communication network through advanced technology to achieve a reliable, safe, economic, and efficient power grid. In addition to ensuring the two-way flow of electric energy among all nodes from the power plant to the end-user in the process of power transmission and distribution, smart grid also needs to transmit a large amount of information; to achieve this, power communication technology is required [1]. The application of power communication technology in smart grid can further improve the efficiency and quality of power systems. Establishing an efficient power communication channel can provide rapid feedback concerning user circumstances, enable effective resource use, and address unexpected power grid situations in a timely manner. The channel can also monitor all abnormal parameters in the power grid at the same time, providing a comprehensive technical guarantee for the energy system.

The key factor in the development of smart grid is certainly the communication infrastructure. PLC and LTE have been used to constitute pathways for implementing smart grid. There are many applications of PLC technology in smart grid communication, especially in transmission networks [2]. In the low-speed application scenario, ultranarrowband PLC technology can realize long-distance communication. Additionally, as a multi-channel transmission system, PLC technology has been developed into a multiple input multiple output (MIMO) version, which uses space-time coding to realize transmission diversity and spatial multiplexing, and can greatly improve the stability and transmission efficiency of the system. PLC technology has become quite mature, and related supporting facilities have become more advanced; hence, PLC technology is widely used in power system communication.

With the promotion of distributed generation (DG), consumers have begun to experience the advantages in generating part of the electricity. Then most commonly used types of power generation are photovoltaic panels and wind turbines. The principle of distributed generation is to achieve the goal of efficient, economic and stable power generation in the distribution network system, through directly setting or focusing on the power generation equipment near the load. As far as reliability is concerned, DG is considered to be a form of diversified power generation and has a variety of methods (a hybrid method) to achieve its objectives. DG can generate electricity not only under normal conditions, but also in some special cases [3]. With the development of LTE and 5G, the application of wireless communication in power system communication is increasing [4]. In the distribution network, the network facilities are poor and there are many users, and it is difficult to meet the requirements by a separate communication method. The relationships in DG can be applied to the data communication infrastructure to combine and create redundancy for critical and necessary data communication paths. This is the so-called heterogeneous data communication architecture or hybrid data communication system [5].

At present, research on wired / wireless heterogeneous communication is still limited, mainly focusing on heterogeneous wireless communication. Yasin Kabalci clearly revealed the network and physical structure of smart grid [6]. T. De Schepper proposed the ORCHESTRA framework to manage different devices in heterogeneous wireless networks and introduced capabilities such as packet-level dynamic and intelligent handovers (both inter- and intra technology), load balancing, replication, and scheduling [7]. Zhang developed a set of hybrid communication system simulation models to verify the key system design standard of distributed solar photovoltaic (PV) communication systems [8]. J. Wan proposed a heterogeneous network architecture based on a software-defined network (SDN) for realizing cross-network flexible forwarding of multi-source manufacturing data and optimized the utilization of network resources [9]. M. A. Zarrabian studied the problem of multirate packet delivery in heterogeneous packet-erasure broadcast networks and presented a new analytical framework for characterizing the delivery rate and delivery delay performance of a nonblock-based network coding scheme [10]. Koohifar proposed a hybrid Wi-Fi/LTE aggregated data communication architecture and compared it with the baseline LTE architecture under the smart grid flow profile. The results show that the proposed architecture can improve the performance of the control channel and random access channel [11].

The first method combined PLC and wireless communication in smart grid data transmission was developed in [12]. However, the article only compared the achievable data rates in communication systems and analyzed the advantages of hybrid systems over non-hybrid systems.

There are many devices in the distribution network, that generate a large amount of data, and the throughput that needs to be transmitted is large. Meanwhile, telemetry and remote signal data have transmission delay requirements. Therefore, we must reasonably allocate data to PLC or LTE while meeting the data delay requirements and reliability to maximize system throughput.

This paper proposes a throughput-optimized transmission scheduling framework that meets the smart grid specifications and system requirements. In addition, we prove that the proposed heterogeneous PLC and LTE communication uplink optimized transmission scheduling methods can maximize system throughput and reasonably allocate resources.

Our contributions are as follows:

(1) We propose a framework for heterogeneous communication in smart grid that combines PLC and LTE uplink channels to maximize the throughput of the system.

(2) We prove the stability and solvability of the scheduling system in theory and design an optimization algorithm to solve the scheduling problem of system throughput optimization.

(3) The simulation results show that the optimized transmission scheduling method for heterogeneous PLC and LTE communication can maximize the throughput of the system compared with other typical solutions.

The remainder of this paper is organized as follows. In section II a network framework for PLC and LTE communication in smart grid is designed. In section III, the stability and solution of the system are theoretically proven by reasoning. In section IV, the simulation results show that the optimized transmission scheduling method of heterogeneous PLC and LTE communication can maximize the throughput of the system. Finally, we conclude this paper in section V.

## 2 System model

First, consider the system application scenario shown in Figure 1. As shown in the figure, the system consists of three parts: N wireless sensor network nodes with different types of sensors, intelligent grid multi interface heterogeneous communication platforms, and remote cloud monitoring equipment. There are three modules in the multi-interface heterogeneous communication platform of the smart grid: the scheduler, PLC module and LTE module. First, different types of sensors collect all kinds of data, including voltage, current, temperature and humidity, control and other data, and then transmit the data to the multi-interface heterogeneous communication platform of smart grid. There are different transmission delay and packet loss rate requirements for the different types of sensors, and these requirements are comprehensively analyzed to achieve the maximum system throughput and to decide whether to use the PLC or LTE of the data.

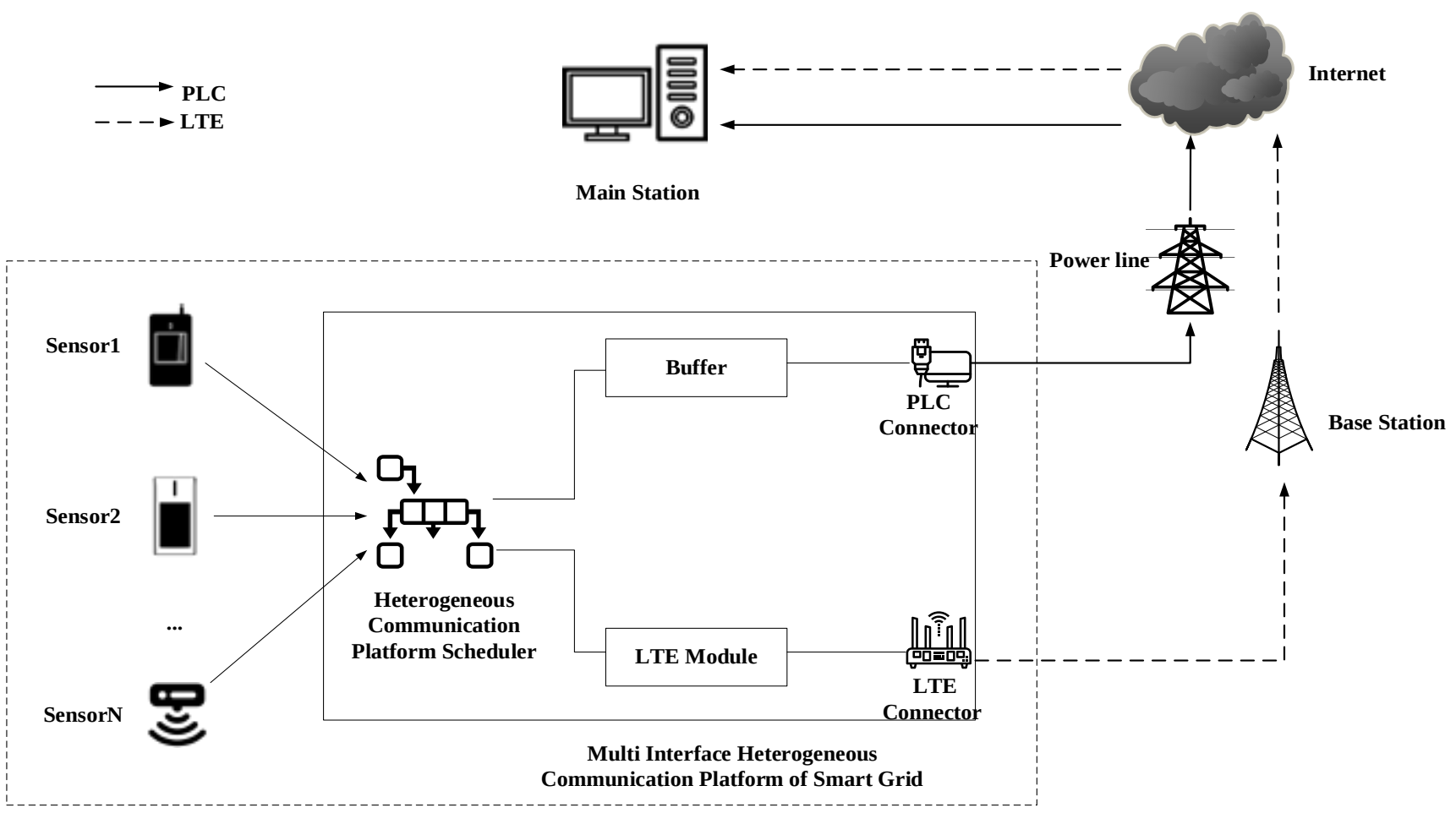

**Fig. 1.** System structure diagram

These three modules work together to optimize the transmission, which means that to maximize the throughput of the system, the data sampled by different sensors are reasonably distributed to PLC and LTE. Specifically, the workflow of the system is as follows: all kinds of data collected by N different types of sensors are transmitted to the scheduler in the heterogeneous communication platform. It comprehensively analyzes the specific transmission delay and data packet loss rate requirements for each piece of data and reasonably distributes the data to the PLC or LTE. If the data is transmitted by PLC, it will pass through the data buffer, and then it will be transmitted to the remote cloud monitoring system by wire through the PLC interface. If the LTE uplink channel interface is selected, we do not consider the delay, but due to the characteristics of the wireless channel, packet loss may occur.

There is a data buffer in the PLC module, and it is used to temporarily store the transmitted data. When the data are transferred to the front end of the data buffer, the end of the data buffer is simultaneously transferring data to the outside. In the buffer, the data are transmitted in line, according to the principle of "first in, first out", so the data will stay in the data buffer for a period of time. The data buffer will affect the transmission delay of the data. We assume that the transmission data rate $r_{PLC}$ is constant. The total time for data transmission in PLC cannot exceed the system delay requirements.

In the LTE part, the data transmission mode is the space-time multiplexing mode of orthogonal frequency division multiple access (OFDMA), so the transmitted data blocks occupy a certain range on the time axis and on the space axis. As shown in Figure 2, generally, a resource block (RB) occupies 0.5 ms, which occupies 12 subcarrier channels. We assume that the wireless transmission resource block is limited for a period of time, and the resource block consumed by wireless transmission data during resource allocation cannot exceed this fixed value. At the same time, packet loss of data will occur during wireless transmission, resulting in data transmission failure. Bit error rate $P_b$ during wireless transmission is related to the signal-to-noise ratio threshold and the average signal-to-noise ratio of the platform. Different types of data have different requirements for successful reception rates, and the probability of successful data transmission cannot be less than the corresponding requirements for successful reception rates.

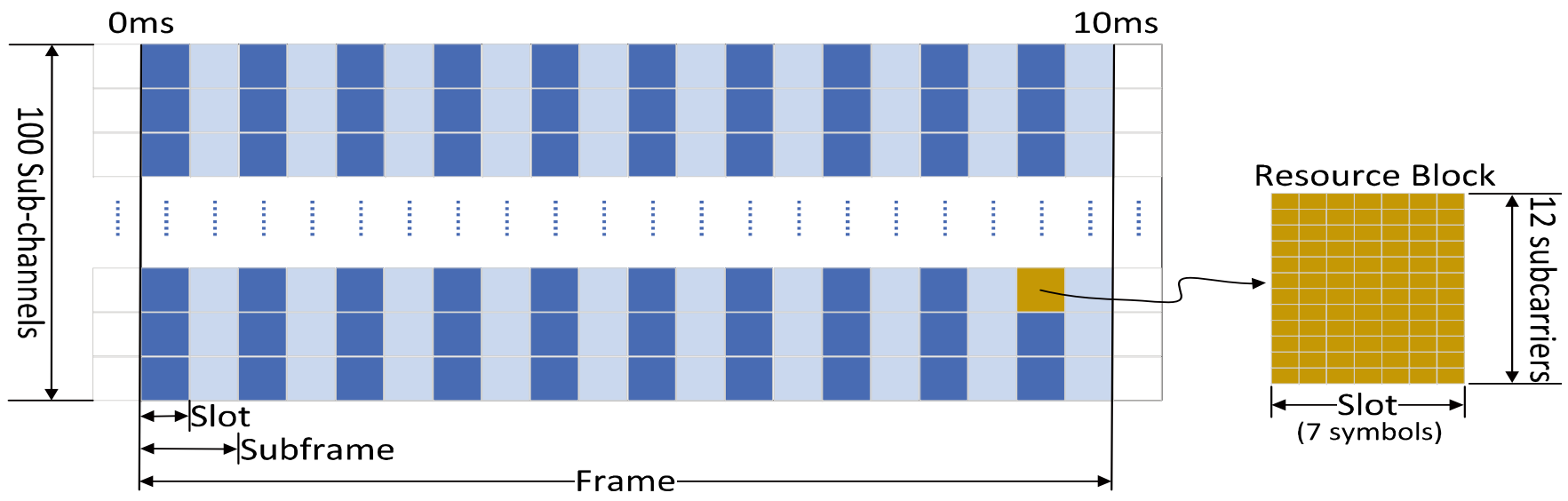

**Fig. 2.** Schematic diagram of wireless resource blocks

## 2.1 Problem formulation

### a. Dynamics of PLC data buffer

In the process of wired transmission, the total time needed for data transmission is equal to the total time spent in the data buffer plus the time needed for wired transmission. Let $D_{n,i}$ represent the data size sampled by the $n$th sensor at the $i$th timestep; $t_{n,i}$ represents the total time that the amount of data $D_{n,i}$ sampled by the $n$th sensor at the $i$th timestep stays in the data buffer during wired transmission. The total time in which the data remain in the data buffer is the ratio of the current buffer depth to the wire transmission rate.

$$t_{n,i} = \frac{m(i)}{r_{PLC}} \tag{1}$$

The depth of the current data buffer is equal to the depth of the data buffer at the previous moment plus the amount of incoming data minus the amount of data transmitted. We use $m(i)$ to represent the depth of sampled data amount $D_{n,i}$ in the buffer at time slot $i$; $x_{n,i}$ represents the data $D_{n,i}$ is transmitted by PLC or LTE, i.e., decision variable $x_{n,i} = 1$ represents data transmitted by PLC and $x_{n,i} = 0$ represents data transmitted by LTE.

$$m(i+1) = m(i) + \sum_{n=1}^{N} \left( D_{n,i} \cdot x_{n,i} - r_{PLC} \cdot x_{n,i} \right) \quad (2)$$

**b. Bit error rate of wireless transmission**

The bit error rate is an index used to measure the accuracy of data transmission in a given time, and the generation of the error code is due to the decay in the signal voltage in the signal transmission, which causes the signal to be damaged in the transmission. In [13], equation (3) is used to determine bit error rate $P_b$ during wireless transmission:

$$P_b = \frac{1}{2}\left[1 + \operatorname{erf}\left(\frac{\gamma - \bar{\gamma}}{\sigma\sqrt{2}}\right)\right] \quad (3)$$

In equation (1), $\sigma$ is a fixed parameter, $\gamma$ represents the signal-to-noise ratio threshold, $\operatorname{erf}(\cdot)$ denotes the error function, and $\bar{\gamma}$ is the average signal-to-noise ratio of the sensor. Additionally,

$$\bar{\gamma} = \frac{P_r}{N_0} = \frac{P_t K}{N_0}\left(\frac{d_0}{d_n}\right)^{\lambda} \quad (4)$$

**c. Objective function and constraints**

The throughput of a network system is the measure of the network's bearing capacity and transmission performance. In this paper, we try to maximize the throughput of a heterogeneous platform, which is the sum of the data volume of the PLC and LTE in the infinite time domain and is used as the system's throughput, represented by TP.

$$\max \mathrm{Tp} = \sum_{i=1}^{\infty}\sum_{n=1}^{N} D_{n,i} \cdot x_{n,i} + \sum_{i=1}^{\infty}\sum_{n=1}^{N} D_{n,i} \cdot (1 - x_{n,i}) \cdot (1 - P_b)^{D_{n,i}\cdot(1-x_{n,i})} \quad (5)$$

According to the delay and reliability requirements of various types of smart grid communication data, the following constraints hold.

$$\frac{D_{n,i} \cdot x_{n,i}}{r_{PLC}} + t_{n,i} \le T_{c,n} \quad (6)$$

$$1 - (1 - P_b)^{D_{n,i}\cdot(1-x_{n,i})} \le 1 - P_{e,n} \quad (7)$$

$$\frac{\sum_{i=1}^{T}\sum_{n=1}^{N} D_{n,i}\cdot(1-x_{n,i})}{R_{RB}} \le Y \quad (8)$$

$$0 \le X_{n,i} \le 1, x_{n,i} \in N \quad (9)$$

Constraint (6) indicates that the total time for data transmission by wire cannot exceed delay requirement $T_{c,n}$ of the nth sensor data.

Constraint (7) means that the bit error rate of data transmitted wirelessly cannot exceed the required bit error rate of data $(1 - P_{e,n})$, and $P_{e,n}$ indicates the successful receiving rate of the data transmitted wirelessly for the $n$th sensor.

Constraint (8) means that the number of resource blocks consumed by the data transmitted wirelessly in time $T$ does not exceed the number of wireless transmission resource blocks $Y$, and $R_{RB}$ is the transmission rate of the LTE resource blocks.

Constraint (9) shows that a packet can only be transmitted in either a wired or wireless mode.

In this paper, on the premise of fully considering the delay requirements of the data in wired transmission and the reliability requirements of the data in wireless transmission, we decide to take either 0 or 1 for every $x_{n,i}$, which indicated whether the data of each sensor at each time is transmitted by wire or by wireless, to maximize the total throughput of the system. Obviously, this is a nonlinear 0-1 programming problem with constraints.

## 3 Problem solving

The optimization of the heterogeneous platform throughput in an infinite time problem cannot be solved by calculation. In this paper, we use an iterative method and obtain the optimal solution of the current state and the current status by predicting the optimal solution at time *NP*, which is optimized in the following section.

For the dynamics of the PLC data buffer:

$$m(k+1) = m(k) + \sum_{n=1}^{N} \left( D_{n,k} \cdot x_{n,k} - r_{PLC} \cdot x_{n,k} \right) \tag{10}$$

We mark as:

$$m(k+1) = f(m(k), x(k)) = m(k) + \sum_{n=1}^{N} \left( D_{n,k} \cdot x(k) - r_{PLC} \cdot x(k) \right), m(0) = m_0 \tag{11}$$

where $m(k) \in R$, and $x(k) \in R$ represent the depth of the data buffer and the decision variable at time $k$ ,respectively. The system's decision volume and data buffer constraints are:

$$m(k) \in [0, M], \quad k \geq 0, \tag{12}$$

$$x(k) \in [0,1], \quad k \geq 0, \tag{13}$$

where $M$ is the maximum buffer depth.

Obviously, the mapping $f$ satisfies the following conditions:

A1) $f: R \times R \to R$ is continuous,

A2)$[0,1] \in R$ is a tight interval and a convex interval, $[0, M] \in R$ is connected, and point (0,0) is contained in the aggregate [0,1] × [0, M];

### 3.1 Finite time domain problem reformulation

$$\max \mathrm{Tp} = \sum_{i=0}^{\infty}\sum_{n=1}^{N}\left(D_{n,i}\cdot x(i) + D_{n,i}\cdot(1-x(i))\cdot(1-P_b)^{D_{n,i}\cdot(1-x(i))}\right)$$

*TP* is denoted as

$$TP^{\infty}\left(m(0), x(\cdot)\right) = \sum_{i=1}^{\infty} F(m(i), x(i)) \tag{14}$$

where F(·,·) satisfies the following conditions:

B1)F$(m(i), x)$: $M \times X \rightarrow R$ ,where R is continuous with respect to independent variables $m(i)$ and $x$ and satisfies $\mathrm{F}(0,0) = 0$ while for any $(m(i), x) \in M \times X \backslash \{0,0\}$ , there are $\mathrm{F}(m(i), x) > 0$.

Due to the existence of time-domain constraints, it is almost impossible to give the exact expressions for analytical solutions to problem (14). To solve this problem, we can rewrite (14) as follows:

$$TP^{\infty}\left(m(0), x(\cdot)\right) = \sum_{i=0}^{T-1}\sum_{n=1}^{N}\left(D_{n,i}\cdot x(i) + D_{n,i}\cdot\left(1-x(i)\right)\cdot(1-P_b)^{D_{n,i}\cdot(1-x(i))}\right) + \sum_{i=T}^{\infty}\sum_{n=1}^{N}\left(D_{n,i}\cdot x(i) + D_{n,i}\cdot\left(1-x(i)\right)\cdot(1-P_b)^{D_{n,i}\cdot(1-x(i))}\right) \tag{15}$$

$$\Omega: \;= \{m(i) \in R \mid E(m) \leq \alpha, \alpha > 0\} \tag{16}$$

where here $E(m(i))$ is a positive definite function such that $E(0) = 0$ and $E(m(i)) > 0, \forall m(i) \neq 0$. Assume that such an input $x = \kappa(m)$ in $\Omega$ satisfies the following conditions:

C1) $\Omega \subseteq \mathrm{M}$,

C2) $\kappa(m) \in X$ for all $m \in \Omega$,

C3) For all $x \in \Omega$ positive definite function $E(m)$ satisfies the inequality:

$$E\left(m(s)\right) - E\left(m(j)\right) \leq -\sum_{i=j}^{s}\sum_{n=1}^{N}\left(D_{n,i}\cdot x(i) + D_{n,i}\cdot\left(1-x(i)\right)\cdot(1-P_b)^{D_{n,i}\cdot(1-x(i))}\right), \quad T \geq s \geq j \geq 0 \tag{17}$$

In $\Omega$, because the input $x = \kappa(m)$ makes the data buffer system asymptotically stable, $E\left(m(\infty)\right) = 0$, and

$$\sum_{i=T}^{\infty}\sum_{n=1}^{N}\left(D_{n,i}\cdot x(i) + D_{n,i}\cdot\left(1-x(i)\right)\cdot(1-P_b)^{D_{n,i}\cdot(1-x(i))}\right) \leq E\left(m(T)\right) \tag{18}$$

Substituting the above formula into (15),

$$\sum_{i=0}^{T-1}\sum_{n=1}^{N}\left(D_{n,i}\cdot x(i) + D_{n,i}\cdot\left(1-x(i)\right)\cdot(1-P_b)^{D_{n,i}\cdot(1-x(i))}\right) + \sum_{i=T}^{\infty}\sum_{n=1}^{N}\left(D_{n,i}\cdot x(i) + D_{n,i}\cdot\left(1-x(i)\right)\cdot(1-P_b)^{D_{n,i}\cdot(1-x(i))}\right) \leq \sum_{i=0}^{T-1}\sum_{n=1}^{N}\left(D_{n,i}\cdot x(i) + D_{n,i}\cdot\left(1-x(i)\right)\cdot(1-P_b)^{D_{n,i}\cdot(1-x(i))}\right) + E\left(m(T)\right) \tag{19}$$

Therefore, we define the finite time domain objective function as follows:

$$\sum_{i=0}^{\infty}\sum_{n=1}^{N}\left(D_{\mathrm{n},i}\cdot x(i) + D_{n,i}\cdot\left(1-x(i)\right)\cdot(1-P_b)^{D_{n,i}\cdot(1-x(i))}\right) \coloneqq \sum_{i=0}^{T-1}\sum_{n=1}^{N}\left(D_{n,i}\cdot x(i) + D_{n,i}\cdot\left(1-x(i)\right)\cdot(1-P_b)^{D_{n,i}\cdot(1-x(i))}\right) + E\left(m(T)\right) \tag{20}$$

The above is an upper bound on objective function (14) in the infinite time domain. In this way, we transform the optimization problem of the infinite time domain into a finite time domain optimization problem.

### 3.2 Finite time domain optimization problem

The optimization problem of quasi infinite time domain nonlinear predictive control for discrete systems can be described as

$$max_{\bar{x}(\cdot)}\ TP\left(m(k),\ \bar{x}(\cdot)\right) \tag{21}$$

$$TP\left(m(k),\ \bar{x}(\cdot)\right) = \sum_{i=0}^{T-1}\sum_{n}^{N}\left( D_{n,k+i|k}\cdot\ \bar{x}(k+i|k) + D_{n,k+i|k}\cdot\big(1- \bar{x}(\mathrm{k}+i|k)\big)\cdot(1-P_b)^{D_{n,k+i|k}\cdot(1-\ \bar{x}(\mathrm{k}+i|k))}\right) +\ E\left(\ \bar{m}(k+T|k)\right) \tag{22}$$

This meets both system dynamics and time domain constraints:

$$m(\tau+i|k) = f\left(\ \bar{m}(\tau|k),\ \bar{x}(\tau|k)\right) =\ \bar{m}(\tau|k) + \sum_{n=1}^{N}\left(D_{n,(\tau|k)}\cdot\ \bar{x}(\tau|k) - r_{PLC}\cdot\ \bar{x}((\tau|k)\right),\ \bar{m}(k|k) = m(k) \tag{23}$$

$$x(\tau|k)\in X, \tau\in[k, k+N-1], \tag{24}$$

$$\bar{m}(\tau|k)\in M, \tau\in[k, k+N-1], \tag{25}$$

$$\bar{m}(k+T|k)\in\Omega, \tag{26}$$

where $T$ is the limited prediction time domain and $\bar{m}(\tau|k)$ is the predicted state trajectory of the system starting at $m(k)$ under the action of decision variable $\bar{x}(\cdot)$.

If optimization problem (21) has a solution, record it as

$$X^*(k) = \begin{bmatrix} \bar{x}^*(k|k) \\ \bar{x}^*(k+1|k)^T \\ \cdots \\ \bar{x}^*(k+T-1|k) \end{bmatrix}^T$$

We select $x^*(k|k)$ as the optimal solution of original problem（14）at time step k：

$$x^*(k) =\ \bar{x}^*(k|k) \tag{27}$$

Then：

$$m(k+1) = f\big(m(k), x^*(k)\big) = m(k) + \sum_{n=1}^{N}\left(D_{n,k}\cdot x^*(k) - r_{PLC}\cdot x^*(k)\right) \tag{28}$$

Let us discuss the stability of the prediction data buffer system.

(1) Suppose A1) to A2) and B1) are true,

(2) For nonlinear system (14), there is an input $x = \kappa(m)$, and the domain $\Omega$ of the equilibrium point and the positive definite function $E(m)$ satisfy the conditions C1) ~ C3),

(3) When $k = 0$ , optimization problem (21) has a feasible solution,

Without considering external disturbances and model error:

(a) For any time $k > 0$ , the optimization problem (21) has a feasible solution,

(b) The closed-loop system is asymptotically stable.

Proof: First, the feasibility of the optimization problem (21) at time $k$ means that it is also feasible at time $(k+1)$ . The buffer depth measured at time $k$ is $m(\text{k})$. It is known from the assumption that there is a feasible solution to the optimization problem at this time, which is recorded as

$$X^*(k) = \begin{bmatrix} \bar{x}^*(k|k) \\ \bar{x}^*(k+1|k) \\ \cdots \\ \bar{x}^*(k+T-1|k) \end{bmatrix} \tag{29}$$

It satisfies control constraint (13), and the corresponding state sequence in the interval $[k+1, k+T]$ is

$$m^*(k) = \begin{bmatrix} \bar{m}^*(k+1|k) \\ \bar{m}^*(k+2|k) \\ \cdots \\ \bar{m}^*(k+T-1|k) \\ \bar{m}^*(k+T|k) \end{bmatrix} \tag{30}$$

It satisfies state constraints (12) and terminal constraints (26), that is,

$$\bar{m}^*(k+i|k) \in M, i \in [k+1, k+T], \tag{31}$$

$$\bar{m}^*(k+T|k) \in \Omega \tag{32}$$

According to the principal of iteration, we apply the first element of the open-loop control sequence to the system to obtain buffer depth $m(k+1)$ at (k + 1). Without considering the model error of external interference, we obtain the following:

$$m(k+1) = \bar{m}^*(k+1|k)$$

With $m(k+1)$ as the initial buffer depth, a candidate solution for selecting open-loop optimization problem (21) at time $(k+1)$ is:

$$X(k+1) = \begin{bmatrix} \bar{x}^*(k+1|k) \\ \cdots \\ \bar{x}^*(k+T-1|k) \\ \kappa(\bar{m}^*(k+T|k)) \end{bmatrix} \tag{33}$$

Obviously, the first $(T-1)$ elements of this candidate solution are the last $(T-1)$ elements of optimal solution (29) at time k, and they all satisfy constraint (13). Since $\bar{m}^*(k+T|k) \in \Omega$ and all m in $\Omega$ have $\kappa(m) \in X$, (33) satisfies control constraint (13).

The prediction state sequence corresponding to (33) is

$$M(k+1) = \begin{cases} \bar{m}(k+i|k+1) = \bar{m}^*(k+i|k), & i \in [1,T] \\ \bar{m}(k+T+1|k+1) = f\left(\bar{m}^*(k+T|k), \kappa\left(\bar{m}^*(k+T|k)\right)\right), & i = T+1 \end{cases} \tag{34}$$

From (32), we obtain

$$\bar{m}(k+i|k+1) \in M, i \in [1,T]$$

That is, the first step, $(T-1)$, of the predicted state sequence satisfies the state constraints. The invariance of class $\Omega$ and system $m(k+1) = f(m(k), \kappa(m(k)))$, implies $\bar{m}(k+T+1|k+1) \in \Omega$, that is, the predicted state sequence satisfies state constraints and terminal constraints.

In summary, $X(k+1)$ given by (34) is a feasible solution to the open-loop optimization problem in (21) at time $(k+1)$. Nature (a) is proven.

The stability of the closed-loop system can be proved, but limited to the page length, the specific proof will not be written.

From the recursive formula, we know that equation (2) can take the following form:

$$m(k+1) = m(k) + \sum_{n=1}^{N}\left(D_{n,k} \cdot x_{n,k} - r_{PLC} \cdot x_{n,k}\right) = m(1) + \sum_{k=0}^{\infty}\sum_{n=1}^{N}(D_{n,k} \cdot x_{n,k} - r_{PLC} \cdot x_{n,k}) \tag{35}$$

### 3.3 Integer Solution

The solution variable of the objective function is relaxed into a continuous variable, and the nonlinear integer programming problem is turned into an easy-to-solve nonlinear programming problem. To find the maximum value of a function under certain conditions, we use the Lagrangian multiplier method to solve the function and establish the Lagrangian function of the nonlinear programming problem according to equation (36):

$$\mathrm{L}(x_{n,i}, \lambda) = \mathrm{TP} + \lambda_1 h_1(x_{n,i}) + \lambda_2 h_2(x_{n,i}) + \lambda_3 h_3(x_{n,i}) + \lambda_4 h_4(x_{n,i}) \tag{36}$$

In Equation (36), $\mathrm{L}(x_{n,i}, \lambda)$ represents a Lagrangian function with respect to $x_{n,i}$, Lagrange multipliers $\lambda$, $h_1(x_{n,i})$, $h_2(x_{n,i})$, $h_3(x_{n,i})$ and $h_4(x_{n,i})$ represent the four constraint functions, and $\lambda_1$, $\lambda_2$, $\lambda_3$, $\lambda_4$ represent the Lagrangian multipliers corresponding to the constraint function. The specific forms of $h_1$, $h_2$, $h_3$ and $h_4$ are:

$$h_1(x_{n,i}) = \frac{D_{n,i} \cdot x_{n,i}}{r_{PLC}} + \tau_{n,i} - T_{c,n} \tag{37}$$

$$h_2(x_{n,i}) = P_{e,n} - (1-P_b)^{D_{n,i} \cdot (1-x_{n,i})} \tag{38}$$

$$h_3(x_{n,i}) = \frac{\sum_{n=1}^{N}\sum_{i=1}^{I} D_{n,i}\cdot(1-x_{n,i})}{R_{RB}} - \Upsilon \tag{39}$$

$$h_4(x_{n,i}) = x_{n,i} - 1 \tag{40}$$

The KKT conditions are established according to equations (37) to (40), and optimal solution $x_{relax}$ of the relaxed nonlinear programming problem is obtained by combining the correlated equations of the KKT conditions:

$$\frac{d(L(x_{n,i},\lambda))}{d(x_{n,i})} = \frac{d(\mathrm{TP})}{d(x_{n,i})} + \lambda_1 \frac{d(h_1(x_{n,i}))}{d(x_{n,i})} + \lambda_2 \frac{d(h_2(x_{n,i}))}{d(x_{n,i})} + \lambda_3 \frac{d(h_3(x_{n,i}))}{d(x_{n,i})} + \lambda_4 \frac{d(h_4(x_{n,i}))}{d(x_{n,i})} = 0 \tag{41}$$

$$h_1(x_{n,i}) \le 0, \quad h_2(x_{n,i}) \le 0, \quad h_3(x_{n,i}) \le 0, \quad h_4(x_{n,i}) \le 0 \tag{42}$$

$$\lambda_1, \lambda_2, \lambda_3, \lambda_4 \ge 0 \tag{43}$$

$$\lambda_1 h_1(x_{n,i}) = 0, \lambda_2 h_2(x_{n,i}) = 0, \lambda_3 h_3(x_{n,i}) = 0, \lambda_4 h_4(x_{n,i}) = 0 \tag{44}$$

The branch and bound method is used to obtain the solution variable of the objective function. Either wired or wireless transmission is used for the amount of data according to the solution variable.

We use algorithm 1 to predict the optimal solution that satisfies the 0-1 constraint at 1-T time:

Objective function $\max \mathrm{Tp} = \sum_{n=1}^{N}\sum_{i=1}^{T} D_{n,i}\cdot x_{n,i} + \sum_{n=1}^{N}\sum_{i=1}^{T} D_{n,i}\cdot(1 - x_{n,i})\cdot(1-P_b)^{D_{n,i}\cdot(1-x_{n,i})}$ is taken as the problem P-1;

**Algorithm 1: Branch and bound method**

**Input:** $x_{relax}$: The optimal solution of the relaxation problem with the KKT condition
$Z_{relax}$ : The optimal objective function value of the relaxation problem
$\varepsilon$ : Any value within 0 to 1
**Output:** $x_{0-1}$: The optimal solution of problem P-1 with 0-1 constraints
$Z_{0-1}$: The optimal objective function value corresponding to the optimal 0-1 solution of P-1

**Initialize** k=0, L=0, U $= Z_{relax}$

1. Select any solution $x_j$ that does not meet the 0-1 constraint condition from the optimal solution $x_{relax}$, that is: $x_j \in (0,1)$;

2. IF $0 \le x_j < \varepsilon$, then add constraint condition $x_j = 0$ to problem p-1 to form sub-problem I; otherwise, add constraint condition $x_j = 1$ to problem p-1 to form sub-problem II, where $\varepsilon$ means any value within 0 to 1;

3. k++, continue to find the solution to the relaxation problem of sub-problem Ⅰ or sub-problem Ⅱ, record it as $x_k$, and record the corresponding optimal objective function value as $Z_k$;

4. Find the optimal objective function maximum value U as the new upper bound, that is:

$$\mathrm{U} = \max\{Z_{k'} | k' = 1,2,\cdots,k\}, x_{k'} \in [0,1];$$

5. Then, from the branches that meet the 0-1 condition, find the maximum value of objective function L as the new lower bound:

$$\mathrm{L} = \max\{Z_{k'} | k' = 1,2,\cdots,k\}, x_{k'} \in \{0,1\};$$

6. IF $Z_{k'} < \mathrm{L}$, then cut the corresponding branch;
7. ELSE IF $Z_{k'} > \mathrm{L}$, and does not meet the 0-1 condition, return to step 2;
8. ELSE it means that the optimal objective function value of all branches is equal to the lower bound: $Z_{k'} = \mathrm{L}$,then assign $Z_{k'}$ to $Z_{0-1}$ and $x_{k'}$ to $x_{0-1}$, and use it as optimal solution of the problem p-1,

Then, we use algorithm 2 to find the actual optimal solution that satisfies the 0-1 constraint at time 1-I:

**Algorithm 2: Throughput scheduling optimization algorithm for heterogeneous PLC and LTE communication uplink**

**Input**：$M(:,1)$: Buffer initial depth
**Output:** $X$: The actual optimal solution meeting the 0-1 constraint

**Initialization**： $m_init = 10\text{^}8$，$X = [0; 0; 0]$
1. The buffer initial depth is $M(:,1)$， let $M(:,1) = m_init$;
2. *FOR* $i$=1→I do;
3. $mk = M(:,i)$;
4. Let $mk$ be initial state $M(1)$ in buffer state equation (35), and use the Lagrangian multiplier method to find optimal solution $x_{relax}$ of the nonlinear programming problem after time T;
5. Use algorithm 1 to find a predicted optimal solution $x_{0-1}$ that meets the 0-1 constraint;
6. Make optimal solution $xk$ at the current moment equal to optimal solution $x_{0-1}(1)$ at the predicted first moment, that is, $xk = x_{0-1}(1)$;
7. $X(:,i) = xk$;
8. Substitute $X(:,i)$ as the control amount into formula (2) to obtain the depth $M(:,i+1)$ of the buffer at the next moment;
9. $mk = M(:,i+1)$;
10. END

# 4 Simulation and experiment

In this section, we will evaluate the performance of our scheduling algorithm. A heterogeneous communication network is considered to be composed of $N$ different types of sensors, one smart grid multiinterface heterogeneous communication platform and one remote cloud monitoring device. Table 1 lists the detailed parameters used to evaluate the network performance. We assume that PLC is not interfered with by the external environment and that the transmission rate, $r_{PLC}$, is constant. At the same time, we assume that the bit error rate of LTE can be calculated by calculating the average signal-to-noise ratio of the sensor.

**Table 1.** Simulation parameters

| Parameters | Values |
|---|---|
| Number of RBs | 100 |
| Initial buffer depth | $10^4$ bytes |
| SNR threshold | 11.5 dB |
| $\sigma$ | 2 |
| Thermal noise variance | 1 dB |
| Maximum transmitting power of sensor | 20 dB |
| Number of packets | 100 |
| Reference distance from system to base station | 1 m |
| Actual distance from system to base station | 10-200 m |
| Pathloss exponent | 2 |
| Bit error rate of LTE | 0.00003%-0.00005% |
| Transmission rate of PLC | $10^4$ bytes/s |
| Resource block transfer rate of LTE | 240 bits/s |

We use d to represent the average amount of data sampled per second by the sensor.

To evaluate our proposed algorithm, we compare the single PLC/LTE algorithm with the greedy algorithm. The single PLC/LTE algorithm assume that there is only one communication mode of PLC or LTE in the communication transmission. Theoretically, the greedy algorithm does not consider global optimization but only makes a local optimal choice. The greedy algorithm determines the optimal wired/wireless resource allocation method to maximize the local throughput. The simulation results are presented in terms of throughput, buffer depth and packet loss rate of different *d*.

Fig.3, Fig.4 and Fig.5 show the system throughput of our algorithm at different times. Greedy is the greedy algorithm, Single LTE represents the situation when LTE transmission is considered separately. We can see that the throughput of the system increases with time. The throughput of our algorithm is the largest, the greedy algorithm is second, and the corresponding throughput of single LTE is the smallest.

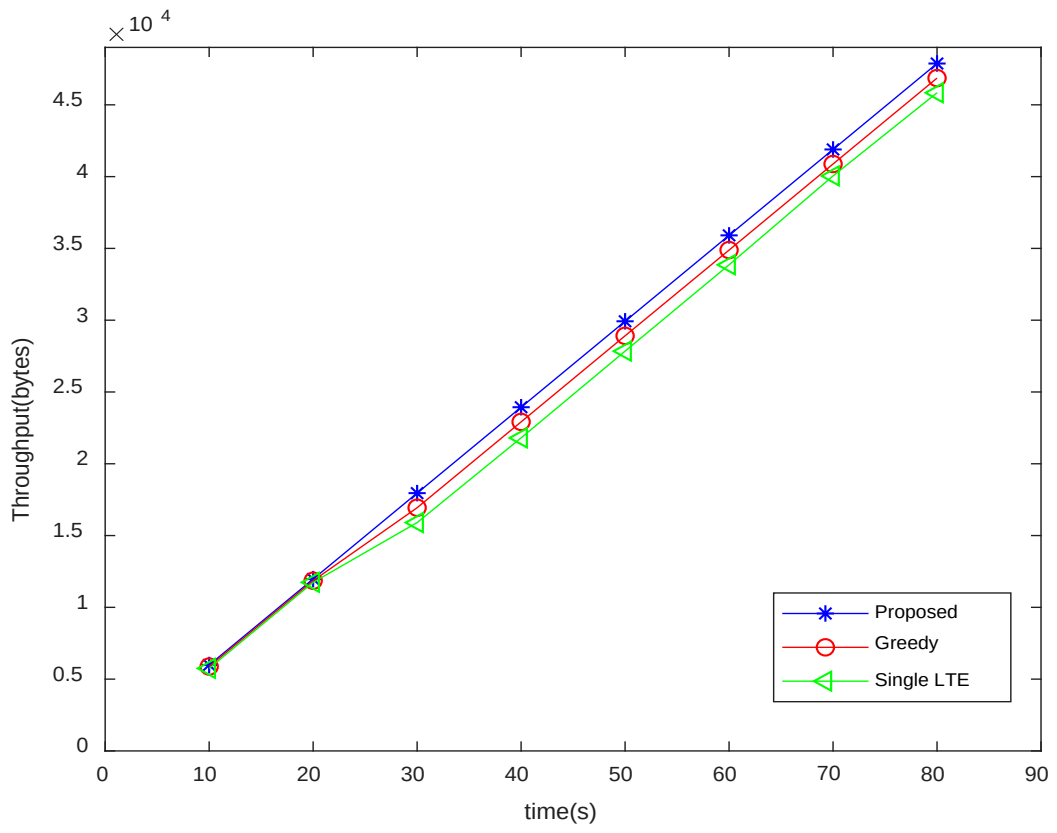

**Fig. 3.** Throughput (d=200) at different times

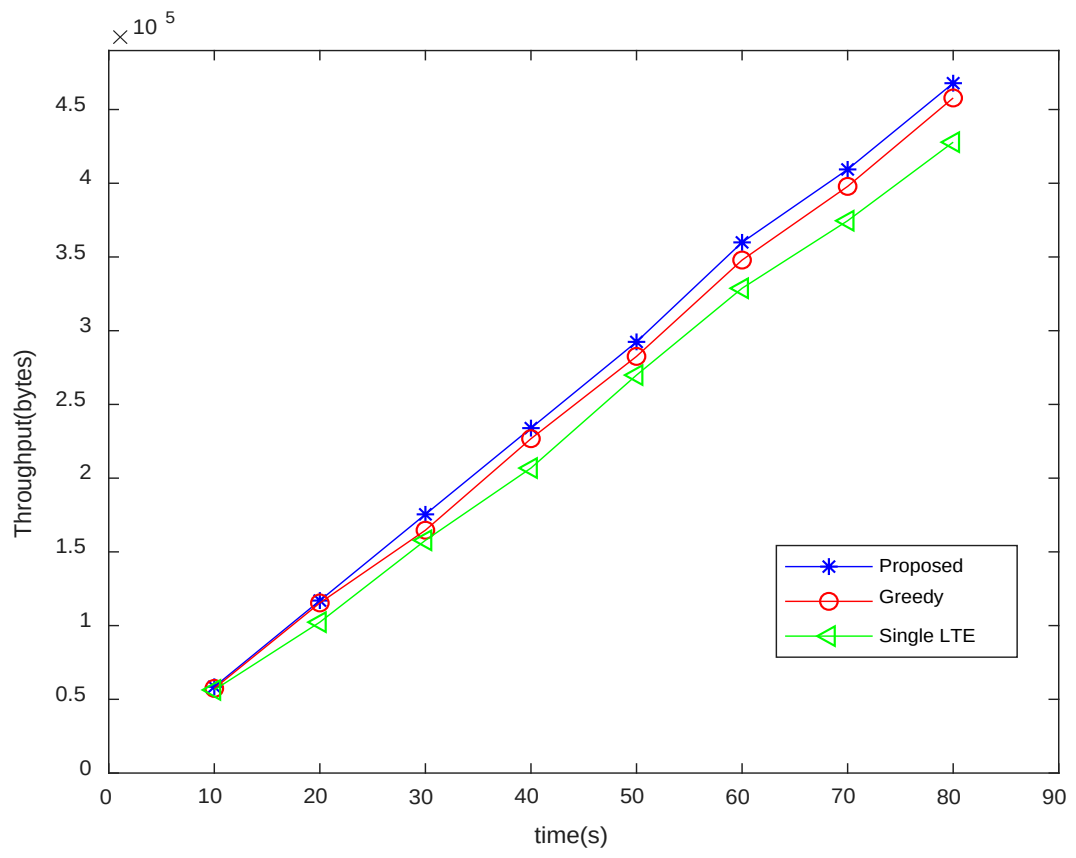

**Fig. 4.** Throughput (d=2000) at different times

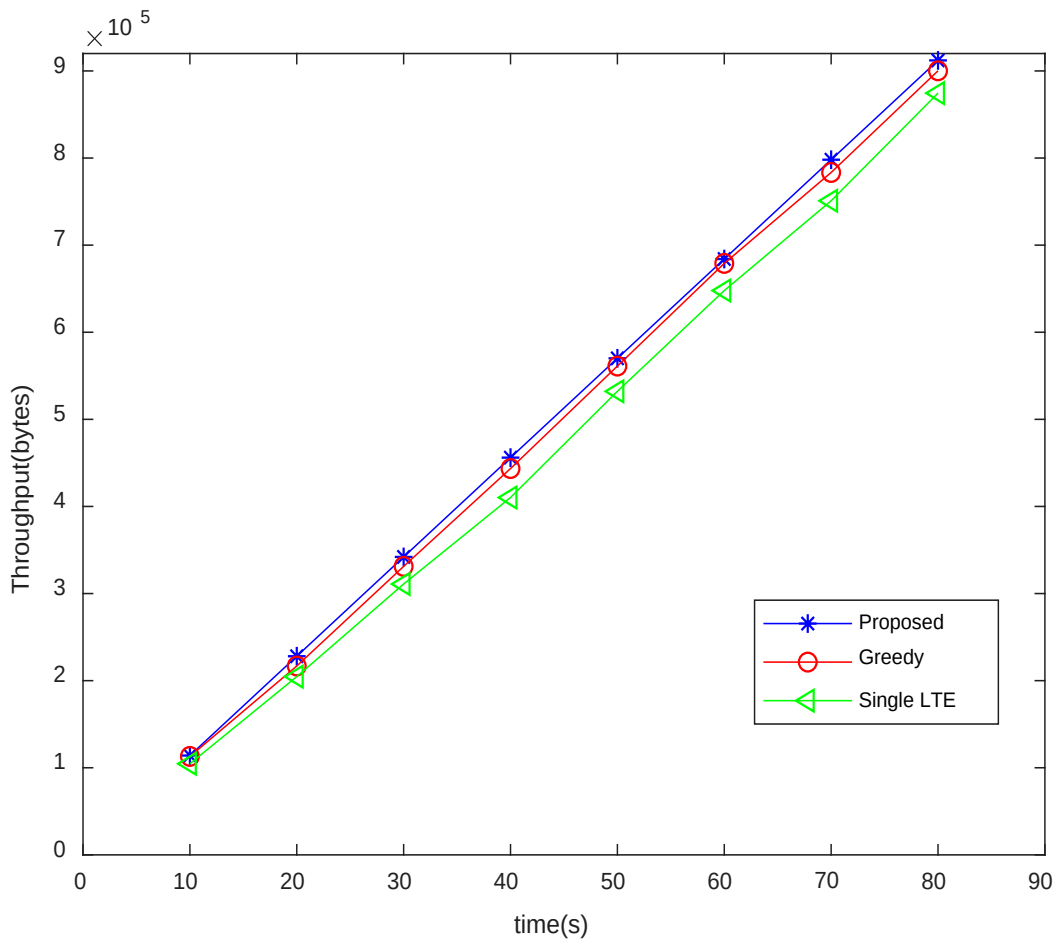

**Fig. 5.** Throughput (d=4000) at different times

Fig.6, Fig.7 and Fig.8 show that the buffer depth of our algorithm at different times is more stable than that of the greedy algorithm and single PLC. Greedy is the greedy algorithm, Single PLC represents the situation when PLC transmission is considered separately. First, we can see that with increasing time, the buffer depth of the greedy method, single PLC and our proposed method decreases. The buffer depth of the proposed method decreases slowly, that of the greedy method decreases slowly at first and then quickly, and that of the single PLC decreases quickly. In a certain period of time, the wireless resources are limited. As time goes on, the wireless resources are gradually insufficient. The greedy algorithm can only allocate most communication resources to PLC, so the buffer depth drops rapidly in about 50 seconds. As time increases, the amount of data collected by the sensors increases. Due to the high wire transmission

rate, the buffer depth will be reduced. However, our algorithm can schedule the sampled data effectively and allocate the resources to PLC or LTE optimally. Therefore, our algorithm makes the buffer depth change more stable.

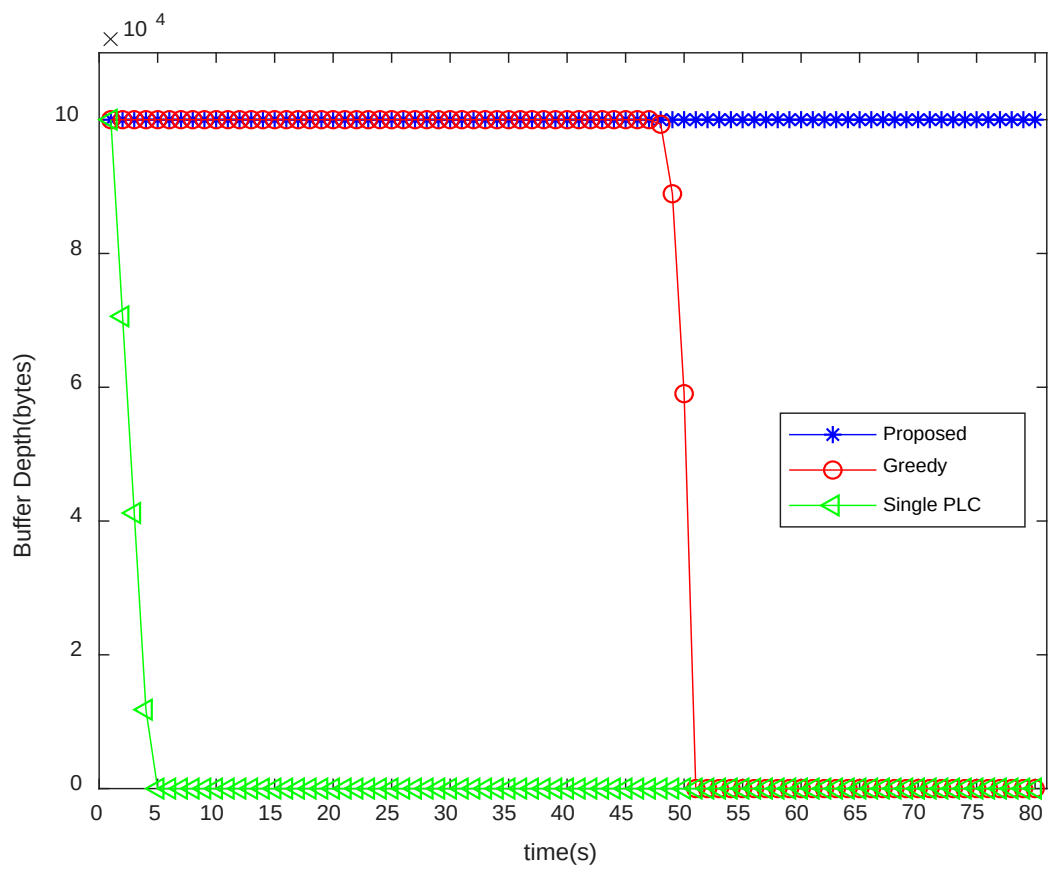

**Fig. 6.** Buffer depth (d=200) at different times

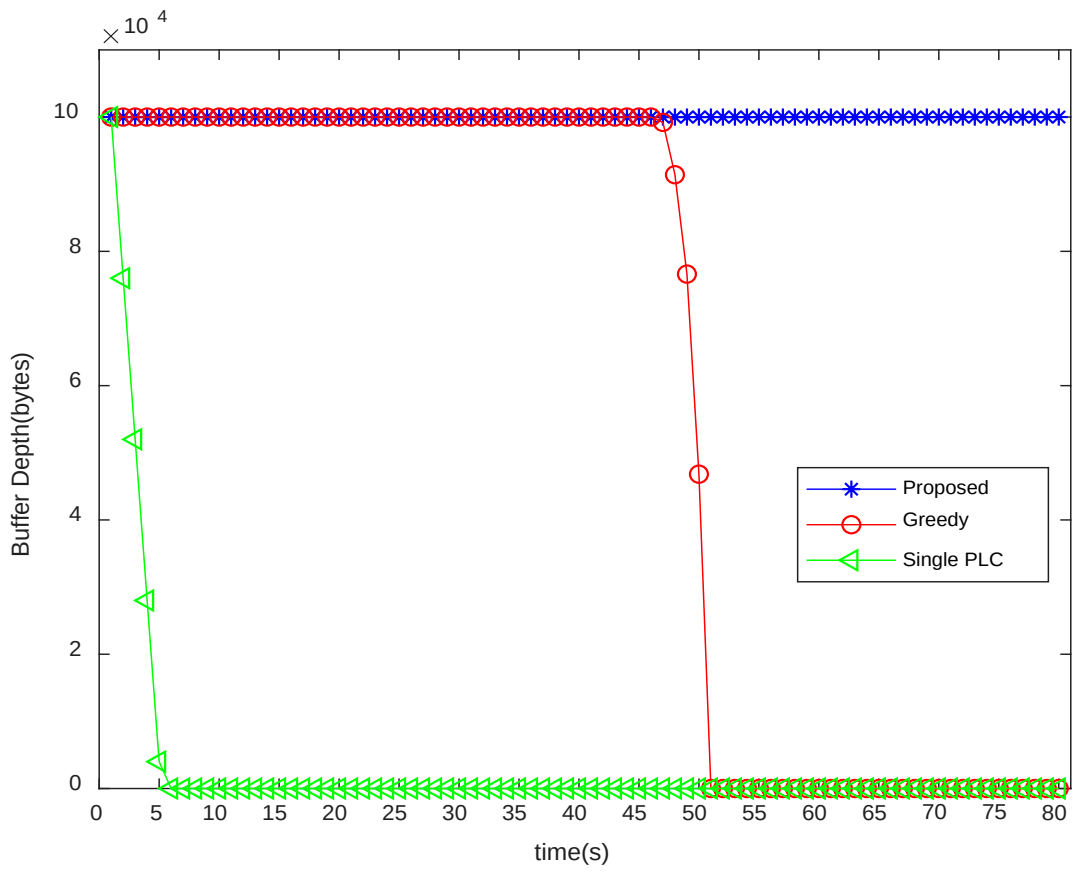

**Fig. 7.** Buffer depth (d=2000) at different times

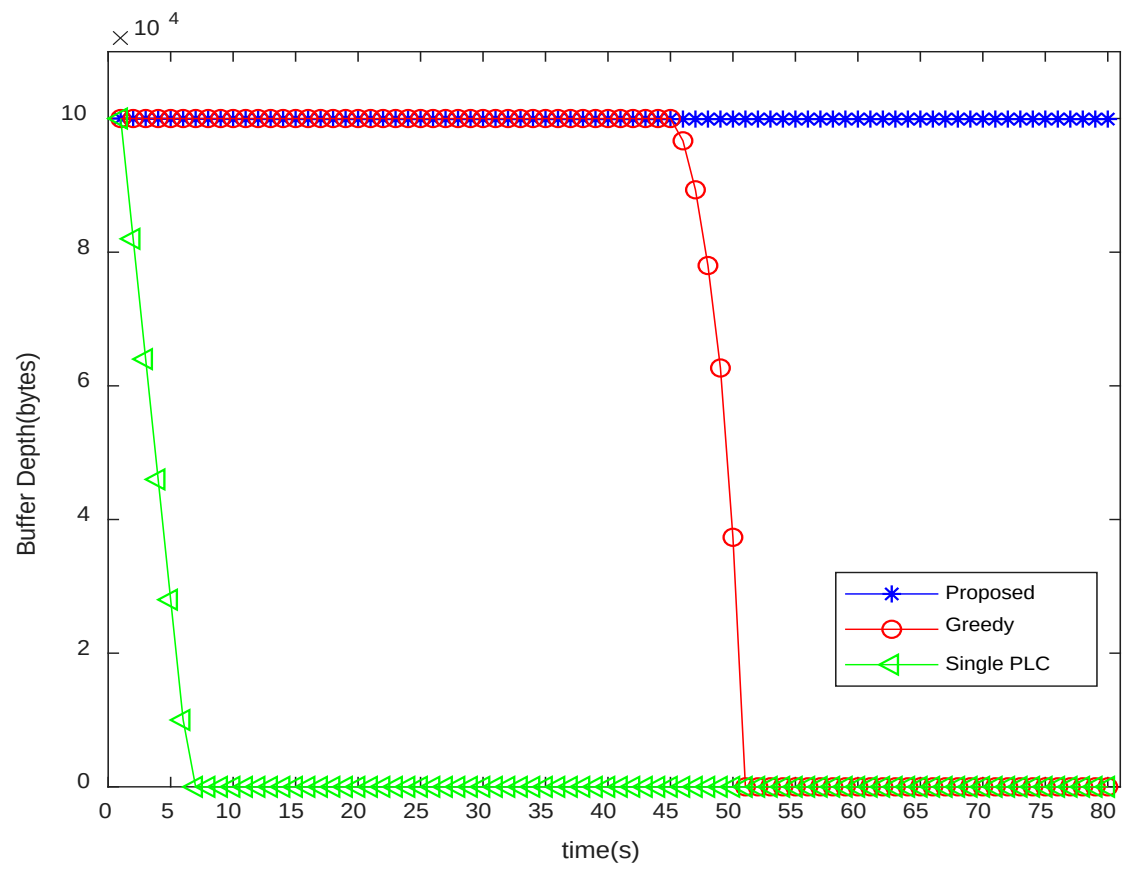

**Fig. 8.** Buffer depth (d=4000) at different times

Fig.9, Fig.10 and Fig.11 show the packet loss rate of our algorithm and the greedy algorithm. Obviously, as the sampling time increases, the packet loss rate of a single LTE is constant, and the packet loss rate of both the proposed algorithms and greedy algorithms increases slowly, but our algorithm can reasonably schedule data to the PLC to reduce the packet loss rate. Our algorithm can achieve better results in terms of packet loss rate.

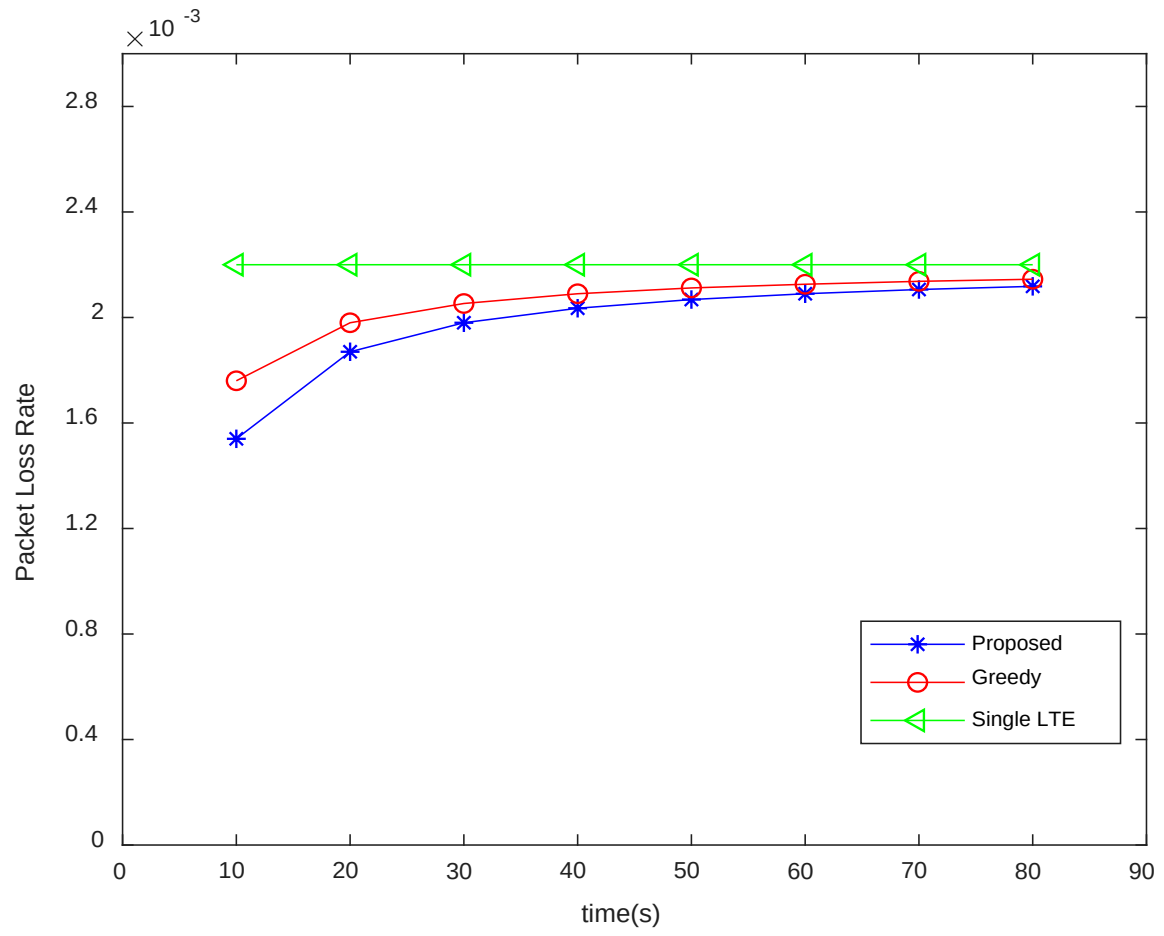

**Fig. 9.** Packet loss rate (d=200) at different sampling times

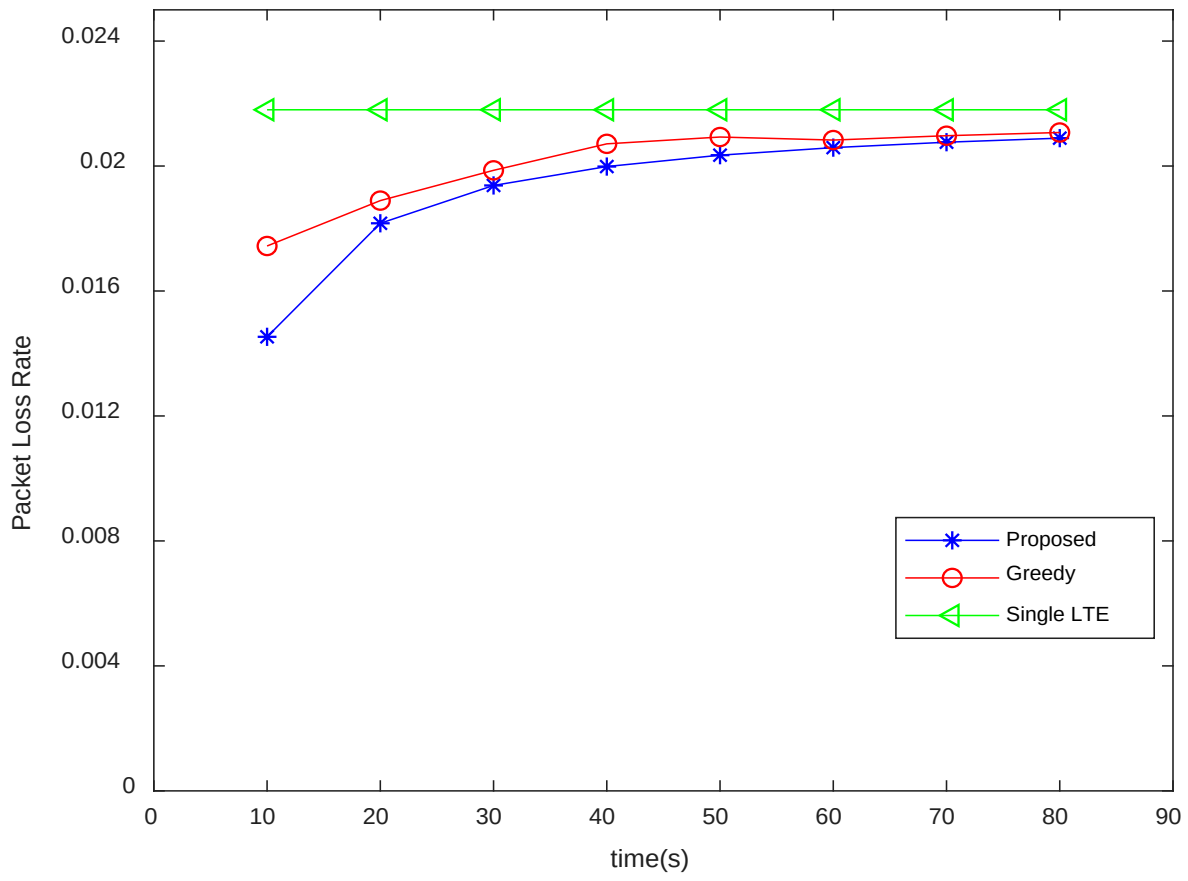

**Fig. 10.** Packet loss rate (d=2000) at different sampling times

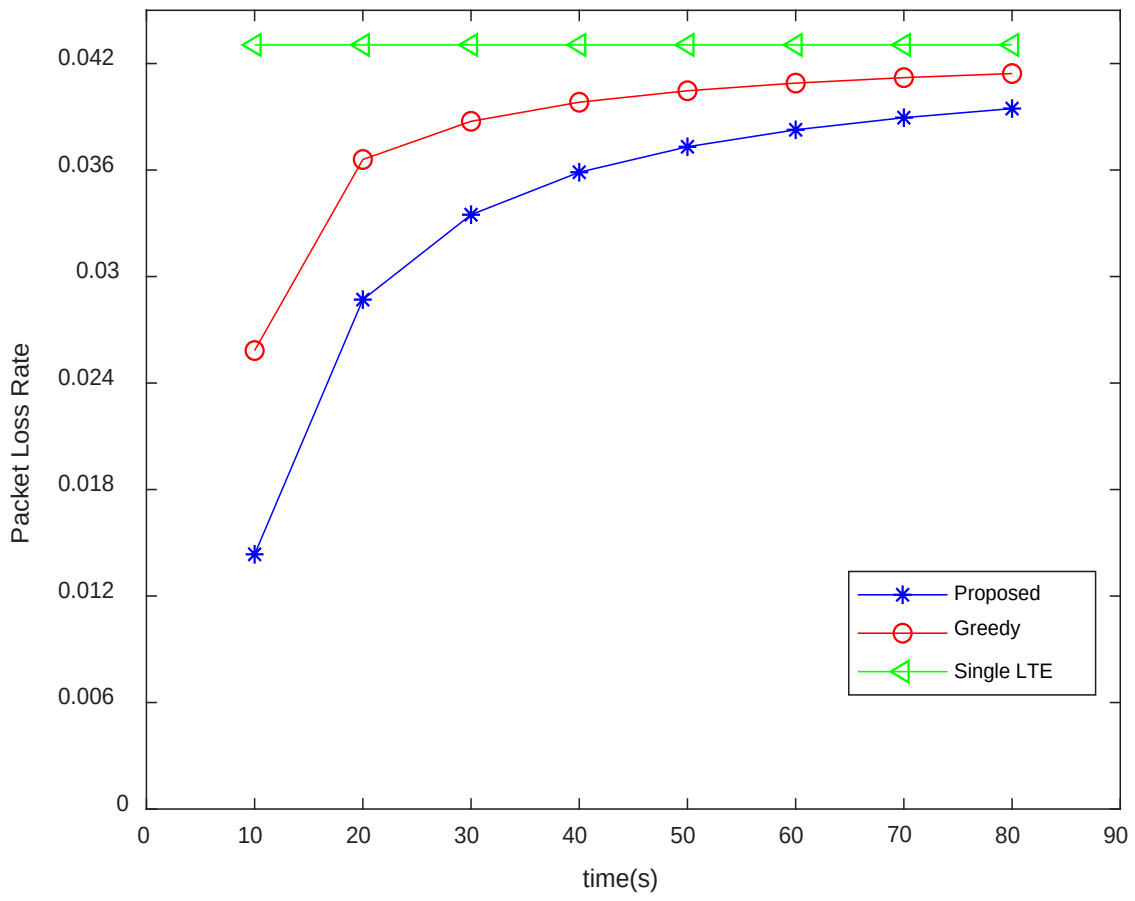

**Fig. 11.** Packet loss rate (d=4000) at different sampling times

# 5 Conclusion

At present, heterogeneous network communication is a difficult problem. Many studies have focused only on wired or wireless communications, and even the research on heterogeneous networks has only focused on how to achieve heterogeneous wireless network communication, there is little research on heterogeneous PLC / LTE. In this paper, we propose a heterogeneous PLC/LTE uplink transmission scheduling method, and establish a heterogeneous communication network system. Then we convert the alloca-tion scheme into a nonlinear programming problem by choosing PLC or LTE to maximize the system throughput. We also prove that the nonlinear problem is solvable and use the branch and bound method to find a 0-1 integer solution. Finally, we compared the performance of our method with other typical solutions. The numerical experimental results show that our algorithm can stabilize the buffer depth and achieve a lower packet loss rate while maintaining maximum system throughput.